\documentclass[aps.pre,showpacs]{revtex4}
\usepackage{graphicx}
\usepackage{color}
\usepackage{ulem}
\def\be{\begin{equation}}
\def\ee{\end{equation}}     
\def\bfi{\begin{figure}}
\def\efi{\end{figure}}
\def\bea{\begin{eqnarray}}
\def\eea{\end{eqnarray}}

\begin{document}

\title{Coarsening and percolation in a disordered ferromagnet}

\author{Federico Corberi$^{1,2}$, Leticia F. Cugliandolo$^3$,
Ferdinando Insalata$^{1,3}$, Marco Picco$^{3}$.}

\affiliation{$^1$ Dipartimento di Fisica ``E.~R. Caianiello'',Universit\`a  di Salerno, 
via Giovanni Paolo II 132, 84084 Fisciano (SA), Italy.
\\
$^2$ INFN, 
Gruppo Collegato di Salerno, and CNISM, Unit\`a di Salerno,Universit\`a  di Salerno, 
via Giovanni Paolo II 132, 84084 Fisciano (SA), Italy.
\\
$^3$ Sorbonne Universit\'es, 
Universit\'e Pierre et Marie Curie - Paris 6, Laboratoire de Physique
Th\'eorique et Hautes Energies, \\
4, Place Jussieu, Tour 13, 5\`eme \'etage,
75252 Paris Cedex 05, France.
}


\begin{abstract}
By studying numerically the phase-ordering kinetics of a two-dimensional ferromagnetic Ising model with quenched disorder 
-- either random bonds or random fields -- 
we show that a critical percolation structure forms in an early stage and is then progressively compactified by the ensuing coarsening 
process. Results are compared with the non-disordered case, where a similar phenomenon is observed, and interpreted within a dynamical 
scaling framework.
\end{abstract}

\maketitle

\section{Introduction} \label{intro}

The phase ordering kinetics following an abrupt change 
of a control parameter, as for instance when a magnetic system is
quenched from above to below the critical temperature, 
is, probably, the simplest non-equilibrium phenomenon in which
interactions among elementary constituents -- spins in the previous
example -- drive the system
from disordered configurations towards an ordered state.
The main feature of such evolution is the coarsening kinetics
whereby initially small domains of the different equilibrium phases
compete and their typical size, $R(t)$, increases in order to reduce 
surface tension. 

Quite generally, this process is endowed with
a dynamical scaling symmetry~\cite{Bray94,Puri-Book} that physically amounts to 
the statistical self-similarity of configurations visited at different times,
so that the only effect of the kinetics is to
increase the size of the domains preserving their morphology. 
Using the terminology of magnetic systems,
for the equal time spin-spin correlation function 
$G(r,t)=\langle s_i(t)s_j(t)\rangle$ with $r\equiv |\vec r_i - \vec r_j|$, scaling implies
\be
G(r,t)={\cal G}\left (\frac{r}{R(t)}\right ).
\label{scal}
\ee
This scaling form is 
expected to apply at distances $r$ much larger than some reference length $r_0$ that is 
usually associated to the lattice spacing, and for times longer than some time $t_0$ that 
is supposed to end a non-universal transient regime where
scaling does not hold yet. Notice that, in principle, Eq. (\ref{scal}) should
be more properly written as
\be
G(r,t)=\widetilde {\cal G}\left (\frac{r}{R(t)},\frac{L}{R(t)}\right ),
\label{scal2}
\ee
where the second entry regulates the presence of finite-size effects when
the domains' size grows comparable to the system's linear dimension $L$. However,
one is usually concerned with the very large system size limit
$R(t)\ll L$ in which the second entry in $\widetilde {\cal G}$
diverges, 
$\widetilde {\cal G}\left (\frac{r}{R(t)},\frac{L}{R(t)}\right )\simeq
\widetilde {\cal G}\left (\frac{r}{R(t)}, \infty \right )$, and Eq. (\ref{scal}) holds
with $\widetilde {\cal G}\left (\frac{r}{R(t)},\infty \right )=
{\cal G}\left (\frac{r}{R(t)}\right )$.
The scaling function ${\cal G}$
goes rapidly to zero for $r>R(t)$, meaning that correlations
have only established up to distances comparable with the typical domain size. 
Regions occupied by the various phases grow similarly in the coarsening
stage, therefore none of them prevails at any finite time if the
thermodynamic limit is taken from the onset. For a magnetic system this
implies that no magnetisation is developed.
The discussion above
applies both to clean systems and to those with quenched disorder, provided it is sufficiently weak not to destroy
the ordered phase or to introduce frustration.

It was shown in a number of papers \cite{Arenzon07,Sicilia07,Sicilia09,Barros09,Olejarz12,Blanchard13,Blanchard14} that, 
starting from a certain time $t_p$, 
some geometrical features of the growing domains in a 
clean two-dimensional coarsening spin system
are very accurately described by the well known properties of
random (site) percolation~\cite{Stauffer,Christensen,Saberi} at the critical threshold $p_c$, where a fractal
spanning or wrapping cluster exists.
This fact comes as a surprise, since percolation is 
a paradigm of a non-interacting problem, whereas coarsening
embodies exactly the opposite. However, although apparently paradoxical,
the interactions among the spins provide the mechanism whereby percolative 
features are built at times of order $t_p$. Indeed, being  the initial state uncorrelated, it is an instance
of random percolation, and the absence of magnetisation sets the concentration 
of up and down spins to $1/2$. Clusters in such a state, therefore, are
too thin to extend across the entire system, 
because $1/2<p_c\simeq 0.59$ on the square lattice used in the simulation. 
Hence, the formation of a spanning or wrapping object at
$t_p$ is certainly an effect of the ordering kinetics.
Quite intriguingly, in fact, as correlations stretch to larger and
larger distances, the droplet structure manages 
to connect initially disjoint parts until 
the largest domain crosses the entire system and acquires the geometric properties of the percolation cluster at threshold.
If $t\ge t_p$ such a domain is surely stable and persists up to 
the longest times (although, obviously, some of its geometrical
features change in time, as we will further discuss).
This is indeed how $t_p$ is defined.

From $t=t_p$ onward, phase-ordering exhibits   
the fractal properties of 
uncorrelated critical percolation on sufficiently large 
scales. As a matter of fact, since the magnetisation remains
zero at any time, such properties occur in a 
correlated system with $p=1/2<p_c$. These two apparently
contradicting instances -- uncorrelated percolation akin to the one occurring at $p=p_c$
on the one hand and correlated coarsening with $p=1/2$ on the other hand --
can be actually reconciled because
they occur on well separated distances:
Percolative properties are confined on length scales
$r \gg R(t)$ larger than those where, due to the interactions,  
some correlation has already developed. 
Instead, on length scales $r< R(t)$, where
the interaction is at work, the 
correlations compactify the fractal domains and the percolative geometry is lost. 
Interactions, therefore, play the two contrasting
roles of promoting the formation of the critical percolation
cluster and of progressively destroying it at scales of order $R(t)$.

This whole framework is quite clearly exhibited in the clean
(i.e.,~non disordered) two-dimensional kinetic Ising model with single spin-flip dynamics
for which $t_p$ was shown to scale with the system size $L$ as~\cite{Blanchard14} 
\be
t_p \propto L^{z_p}
\label{tp}
\ee
where $z_p$ is a new dynamical exponent that depends on the properties of the
lattice populated by the spins. This relation suggests the introduction of a growing length 
associated with the approach to critical percolation
\begin{equation}
  R_p(t) \simeq t^{1/z_p}
  \label{eqrp}
\end{equation}
that saturates reaching the linear system size $L$ at $t=t_p$, namely $R_p(t_p)=L$.
Furthermore, for such non-conserved scalar 
order parameter dynamics, $R(t) \simeq t^{1/z_d}$ with $z_d=2$
quite generally.
Then
\begin{equation}
R_p(t) \simeq (t^{1/z_d})^{z_d/z_p} \simeq  [R(t)]^{z_d/z_p}
\; . 
\label{eq:Rp-Rd}
\end{equation}
On the basis of extensive numerical simulations, it was conjectured
in~\cite{Blanchard14} that $z_d/z_p=n$, the 
coordination of the lattice, for this kind of dynamic rule.
Other clean systems evolving with 
different microscopic rules (local and non-local spin exchange and voter) on various lattices 
are treated in~\cite{Blanchard16}, and more details are given in this manuscript, where the guess
$R_p(t) \simeq [R(t)]^n$ is revisited. Here we 
use Glauber dynamics that we define below. 
Notice that $R_p(t_p)=L$ implies  
\be
R(t_p)=L^{z_p/z_d}.
\label{zpzd}  
\ee

Interestingly, the existence of a percolative structure in coarsening systems is at the heart
of one of the few analytical results in finite dimensional phase-ordering in two dimensions~\cite{Arenzon07,Sicilia07}. In fact,
the distribution of cluster areas in percolation is exactly known \cite{Cardy03} and,
since the evolution of a single hull-enclosed area in a non-conserved scalar order parameter dynamics 
can be inferred,  an exact expression for the area distribution at any time
in the scaling regime can be derived building upon the one at
random percolation. 

As mentioned above, the relevance of percolation in phase-ordering kinetics was
initially addressed in a homogeneous system without quenched disorder.
Real systems, on the other hand, are seldom homogeneous, due to the
occurrence of lattice defects, because of position-dependent external disturbances, or
for other reasons. This fact deeply perturbs the scaling properties \cite{Corberi15c}. Indeed, 
although some kind of symmetry comparable to Eq.~(\ref{scal}) is reported 
in some experiments \cite{Likodimos00,Likodimos01}, the asymptotic growth-law $R(t)$ is observed
to be much slower than in a clean system 
\cite{Likodimos00,Likodimos01,Ikeda90,Schins93,Shenoy99}. Similar conclusions are reached
in the numerical studies of models for disordered ferromagnets, such as the kinetic Ising model
in the presence of random external fields 
\cite{Fisher98,Fisher01,Corberi02,Rao93,Rao93b,Aron08,
Corberi12,Puri93,Oguz90,Oguz94,Paul04,Paul05}, 
with varying coupling constants
\cite{Corberi11,Paul05,Paul07,Henkel06,Henkel08,Oh86,Corberi15,Gyure95},
or in the presence of dilution 
\cite{Corberi15,Puri91,Puri91b,Puri92,Bray91,Biswal96,Paul07,Park10,Corberi13,Castellano98,Corberi15b}.

A natural question, then, is whether the percolation effects observed in bi-dimensional clean systems show up
also in the disordered ones. The answer is not pat because our knowledge in this field
is still rather scarce and even simpler questions regarding
the nature of the dynamical scaling or the form of the 
growth law remain, in the presence of disorder, largely unanswered.  
In~\cite{Sicilia08} the geometric properties of the domain areas in a coarsening 
Ising model with weak quenched disorder were studied
but the analysis of the time needed to reach the critical percolation structure and 
the detailed evolution during the approach to this state were not analyzed.
In this Article we settle this issue. By means of extensive numerical simulations
of the kinetics of both the random field Ising model (RFIM) and the random bond Ising model
(RBIM) we clearly show the existence, starting from a certain time $t_p$, of a 
spanning cluster with the geometry of critical percolation. Not only this 
feature is akin to what was observed in the clean system, but also other quantitative 
properties, such as the size-dependence (\ref{tp}) of the characteristic time $t_p$
and many other details, which all turn out to be exactly reproduced in the presence of disorder
as they are in its absence. This qualifies the relation between coarsening and 
percolation as a robust phenomenon whose origin is deeply rooted into the Physics
of order-disorder transitions in two dimensions.  
Besides interesting and informative on their own, the results of this paper
also represent a step towards a theory of phase-ordering in two-dimensional
disordered ferromagnets, in the same spirit of the one developed~\cite{Arenzon07,Sicilia07}
for clean systems. Needless to say, this would represent an 
important achievement in 
such a largely unsettled field.

This paper is organised as follows.
In Sec. \ref{themodel} we introduce the model that will be considered throughout the paper
and the basic observables that will be computed.
In Sec. \ref{numres} we discuss the outcome of our simulations, both for the clean
case (Sec. \ref{pure}) and the disordered ones (Sec. \ref{disorder}).
We conclude the paper in Sec. \ref{concl} by summarising the results and discussing
some issues that remain open.

 \section{Model and observable quantities} \label{themodel}

We will consider a ferromagnetic system 
described by the Hamiltonian
\be
{\cal H}(\{S_i\})=-\sum _{\langle ij\rangle}J_{ij}S_iS_j+\sum _i H_iS_i,
\label{isham}
\ee
where $S_i=\pm 1$ are Ising spin variables defined on a two-dimensional
$L\times L$ square lattice
(different lattices were considered in~\cite{Blanchard16})
and $\langle ij\rangle$ are two nearest 
neighbor sites. The properties of the coupling constants $J_{ij}$ and
of the external field $H_i$,
specified below, 
define the two disordered models that will be studied in this paper.

{\bf RBIM:} In this case $H_i\equiv 0$ and the coupling constants
are $J_{ij}=J_0+\delta _{ij}$,
where $\delta _{ij}$ are independent random numbers extracted from a flat distribution in 
$[-\delta,+\delta]$, with $\delta <J_0$ in order to keep the interactions
ferromagnetic and avoid frustration effects.

{\bf RFIM:} In this model the ferromagnetic bonds
are fixed $J_{ij}\equiv J_0$ (i.e. $\delta _{ij}\equiv 0$), while the external field $H_i=\pm h$ is
uncorrelated in space and sampled from a symmetric bimodal distribution.

The ordered state, typical of the clean system below the critical temperature
$T_c$, is preserved in the RBIM because the randomness of the coupling constants 
maintains them all ferromagnetic.
Something else occurs in the RFIM, because the external fields efficiently contrast 
the ordering tendency down to $T=0$ in any $d\le 2$.
This is easily explained by the Imry Ma argument \cite{Imry75} which, upon comparing
the energy increase $\Delta E _{surf} \propto J_0\ell ^{d-1}$ due to the reversal of a bubble of size 
${\ell }$ in an ordered phase, to its decrease $\Delta E_{field}\propto h \ell ^{d/2}$
due to the possible alignment of its spins with the majority of the random fields,
shows the existence of a threshold length 
\be
\ell _{IM}\sim \left (\frac{h}{J_0}\right )^{\frac{2}{d-2}}
\label{im}
\ee
above which flipping droplets becomes favorable, thus disordering the system
for $d\le 2$
(it can be shown that ordering is suppressed also right at $d=2$). 
Moreover, still for $d<2$, 
$\ell _{IM} \to \infty$ as $\frac{h}{J_0} \to 0$ so that, if 
such limit is taken from the onset, the system orders in any dimension,
even down to $d=1$ \cite{Corberi02}. In the following, as we will discuss further below, 
we will consider this limit.

Dynamics can be introduced in the Ising model with Hamiltonian (\ref{isham}) 
by flipping single spins according to Glauber transition rates at temperature $T$
\be
w(S_i\to -S_i)=\frac{1}{2}\left [1-S_i\tanh \left (\frac{H^W_i+H_i}{T}\right )\right ]
\label{trate}
\ee
where the local Weiss field
\be
H^W_i=\sum _{j\in  nn(i)}J_{ij}S_j=J_0\sum _{j\in  nn(i)}S_j+\sum _{j\in  nn(i)}\delta_{ij}S_j=
H_{i,det}^W+H_{i,rand}^W,
\ee
is decomposed, for the RBIM,
into a deterministic and a random part (the sum runs over the nearest neighbours 
$nn(i)$ of $i$).

The quenching protocol amounts to evolve a system prepared at time $t=0$ in an
uncorrelated state with zero magnetisation -- representing an equilibrium configuration
at $T=\infty$ -- by means of spin flips regulated by the transition rates (\ref{trate}) evaluated
at the final temperature $T=T_f$ of the quench. 

In this paper we focus on the limit 
$T_f \to 0$ while keeping finite the ratio $\epsilon =\frac{\delta}{T_f}$ (for the RBIM) or
$\epsilon =\frac{h}{T_f}$ (for the RFIM). We will also set $J_0=1$.
The discussion below Eq.~(\ref{im}) points out that in this limit $\ell _{IM}\to \infty$
and phase-ordering always occurs also in the RFIM. 
Moreover,  the transition rates (\ref{trate}) take the simple form
\be
w(S_i\to -S_i)=\left \{ \begin{array}{lcr}
1, & \mbox{for} & H_{i,det}^W S_i <0 \\
0, & \mbox{for} & H_{i,det}^W S_i >0 \\
\frac{1}{2}\left [1-S_i\tanh \left (\frac{H^W_{i,rand}+H_i}{T}\right )\right ], \hspace{1cm}& \mbox{for} & \hspace{.5cm}H_{i,det}^W S_i =0 
\end{array} \right .
\label{simplrates}
\ee
which shows that the model depends only on the ratio $\epsilon$. 
Having a theory with a single parameter is only one of the advantages of
the limit we are considering. The low-temperature limit also reduces thermal noise
and allows for an accelerated updating rule, since Eq.~(\ref{simplrates}) shows that only spins
with $H_{i,det}^W\le 0$ need to be updated. These are, basically, the spins on the corners
of interfaces, a small fraction of the total number of spins in the sample, particularly at long times.

At given times, we have computed the average size of the domains $R(t)$ as the inverse density
of defects \cite{Bray94}, namely by dividing the number $L^2$ of lattice sites by the number of spins 
which are not aligned with all the neighbours. An ensemble average is performed then.

The equal-time pair correlation function, which was defined above Eq.~(\ref{scal}), is computed as
\be
G(r,t)=\frac{1}{4L^2}\sum _{i}\sum _{i_r}\langle S_iS_{i_r}\rangle,
\label{defg}
\ee
where the value of all the spins is measured at time $t$, the $i_r$'s are the four sites at distance
$r$ from $i$ along the horizontal and vertical direction, and $\langle \dots \rangle$ means a 
non-equilibrium ensemble average, that is taken over 
different realisations of the initial configuration and over various
thermal histories, namely the outcomes of the probabilistic updates of the spins.
Let us mention here that we adopt periodic boundary conditions, so
in Eq.~(\ref{defg}) $i_r$ is to be intended {\it modulo} $L$.

As explained in Sec.~\ref{intro}, assuming that $R(t)\ll L$ the two-parameter scaling form
(\ref{scal2}) transforms into the more conventional single-parameter one in Eq.~(\ref{scal}).
However, Eq.~(\ref{eq:Rp-Rd}) shows that another growing length 
$R_p(t)$ is on the scene. The same equation shows also that this quantity grows faster than
$R(t)$ and so, given a finite size $L$ of the system, it may happen that $R_p(t)$ becomes
comparable to $L$ even if the condition $R(t)\ll L$, which is usually considered as the
hallmark of a finite-size free situation, is met. 
In this case a simple scaling form such as the one in Eq.~(\ref{scal}) must be
upgraded to a form akin to Eq. (\ref{scal2}) (with $R_p$ playing the role of $R$ in the
second entry)
in order to take into account the two characteristic lengths.
Indeed, it was shown in \cite{Blanchard14} that a two-parameter scaling
\be
G(r,t)=g\left (\frac{r}{R(t)},\frac{L}{R_p(t)}\right )
\label{scal2par}
\ee
is more appropriate to describe the space-time correlation data.

\begin{figure}[h]
\vspace{2cm}
\centering
\rotatebox{0}{\resizebox{.95\textwidth}{!}{\includegraphics{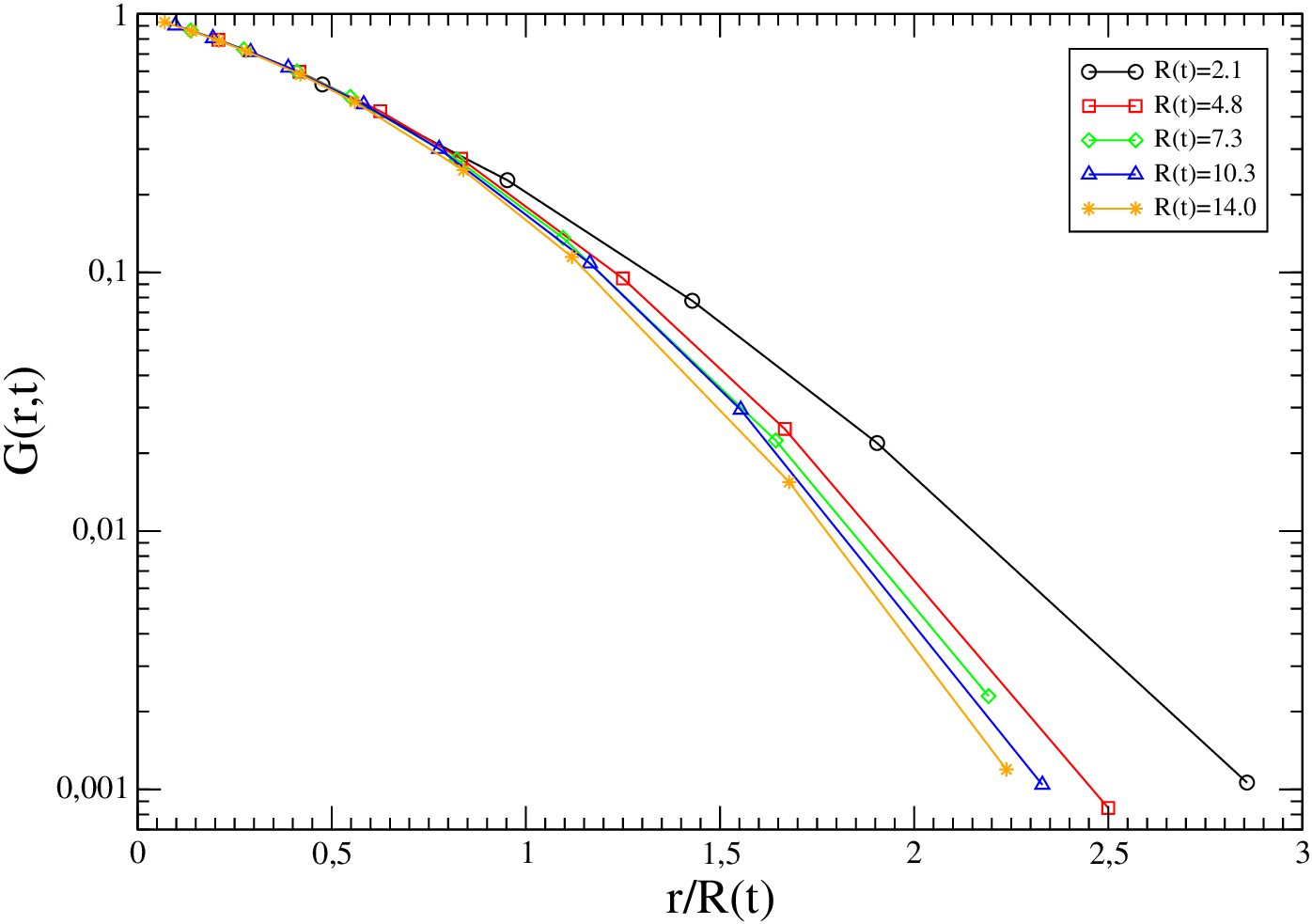}}}
\caption{The spin-spin correlation function $G(r,t)$ is plotted against $r/R(t)$ on a log-linear scale
for a quench of the clean Ising model to $T_f=0$. Different curves correspond to different
values of $R(t)$ (see the key), namely of time $t$. The system size is $L=512$. The deviations from collapse
at large values of the scaling variable $r/R(t)$ are ascribed to the effects of the approach to critical percolation
that has to be taken into account with a second scaling variable, see the text for the explanation of this fact.}  
\label{fig_G}
\end{figure}

The correlation function (\ref{defg}) is shown in Fig. \ref{fig_G} for the clean model quenched to 
$T_f=0$ (more simulation details are given in Sec. \ref{numres}). 
A simple scaling form as in Eq. (\ref{scal}) would imply the collapse of the curves relative to different 
value of $R(t)$ in this kind of plot. Instead one sees here a systematic deviation of the curves
at large $r$ as $R(t)$ increases. As anticipated, this is because the correct scaling form is
Eq. (\ref{scal2par}) instead of Eq. (\ref{scal}). With a two-parameter scaling as in 
Eq. (\ref{scal2par}), plotting against $r/R(t)$ only fixes the first entry, namely $r/R(t)$, but the
second one, namely $L/R_p(t)$ changes as time elapses. This produces the failure of
the collapse observed in Fig. \ref{fig_G}. In order to superimpose the curves one
should keep fixed also the second entry of the scaling function. This can only be done by 
using a different system size $L$ for any different value of $R_p(t)$.
It was shown in \cite{Blanchard14} that in so doing one obtains an excellent scaling in the clean system.
This confirms the validity of the form (\ref{scal2par}).

An important quantity to assess percolative properties is the pair connectedness function,
which in percolation theory is defined as the probability that two points separated by a distance
$r$ belong to the same cluster.
In two dimensions, random percolation theory at $p=p_c$ gives~\cite{Stauffer,Saberi,Christensen}
\be
C_{perc}(r,r_0)\sim \left ( \frac{r}{r_0}\right )^{-2\Delta}
\label{connectperc}
\ee
where $r_0$ is a reference distance, usually the lattice spacing.
Equation~(\ref{connectperc}) holds true for large $r/r_0$, with the critical exponent $\Delta =5/48$. 
The dotted black curve in Fig. \ref{fig_connect} is a numerical evaluation
of $C_{perc}$ on a square lattice with $r_0=1$ and size $L=512$. The curve is
indistinguishable from Eq.~(\ref{connectperc}) except at large distances where
finite-size effects set in. 

In the coarsening system we evaluate this quantity as follows: at a given time, after identifying all the
domains of positive and negative spins, the connectivity is computed as
\be
C(r,t)=\frac{1}{4 L^2}\sum _i \sum _{i_r} \langle \delta _{S_i,S_{i_r}} \rangle,
\label{connect} 
\ee
where $\delta _{S_i,S_j}=1$ if the two spins belong to the same cluster -- namely they are
aligned and there is a path of aligned spins connecting them -- and $\delta _{S_i,S_j}=0$
otherwise.

Another quantity that will be considered is the (average) squared winding angle $\langle \theta ^2\rangle$.
Its definition is the following: At a given time, we chose two points $i,j$ 
on the external perimeter -- the hull --
of a cluster and we compute the winding angle $\theta _{ij} $, namely the angle 
(measured counterclockwise) between the tangent to the perimeter in $i$ and the one in $j$.
After repeating the procedure for all the couples of perimeter points at distance $r$
(measured along the hull), 
taking the square and averaging over the non-equilibrium ensemble, 
one ends up with $\langle \theta ^2(r) \rangle$.
The behavior of $\langle \theta ^2(r)\rangle$ is exactly known in $2d$ critical percolation 
at $p=p_c$~\cite{Duplantier,Wilson}, 
where one has
\be
\langle \theta ^2_{perc}(r,r_0) \rangle =a+\frac{4k}{8+k} \ln \left ( \frac{r}{r_0}\right ),
\label{windexact}
\ee
with $k=6$ and $a$  a non-universal constant.
We will compute this same quantity in the coarsening systems upon
moving along the hulls of the growing domains (in this case, we will only consider the
largest cluster for numerical convenience).

\section{Numerical results} \label{numres}

In this Section we will present the results of our numerical simulations
of the kinetic Ising model. 

We will start  in Sec.~\ref{pure} 
with the clean system since, although 
in this case percolation effects have been already reported \cite{Arenzon07,Sicilia07,Sicilia09,Barros09,Olejarz12,Blanchard13,Blanchard14}, we use 
here tools, such as the connectivity (\ref{connect}), that were not
considered before. This case, therefore, serves
not only to compare with the disordered systems considered further on, but
also as a benchmark for these new quantities. 

\begin{figure}[h]
\centering
\rotatebox{0}{\resizebox{.95\textwidth}{!}{\includegraphics{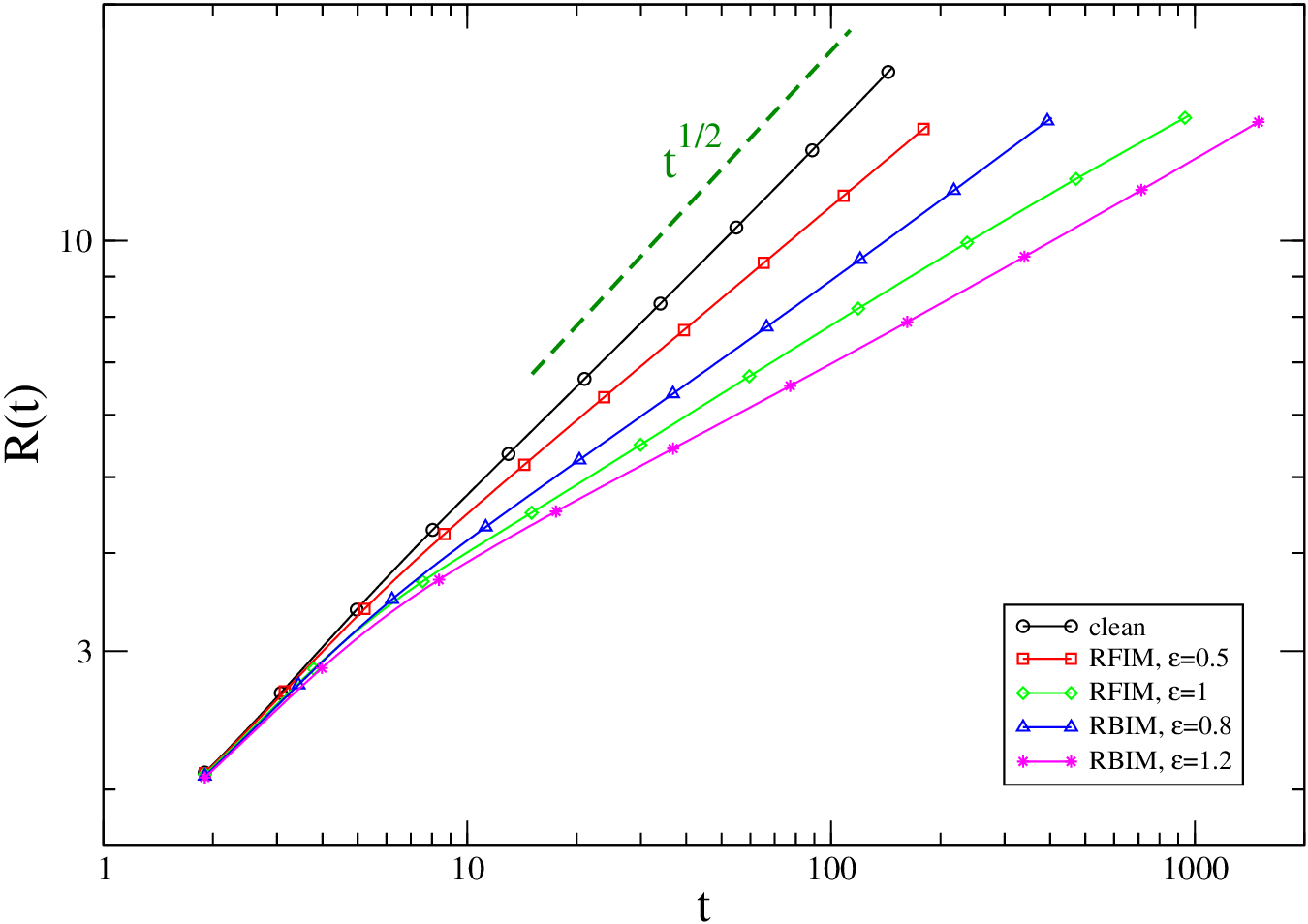}}}
\caption{The growing length $R(t)$ is plotted against time on a log-log scale
for the different models considered in this paper and for different values of
the disorder strength $\epsilon$ (see the key). The dashed green line is the
asymptotic behavior $t^{1/2}$ expected for the clean case.}  
\label{fig_rt}
\end{figure}

We will then turn to the behavior of the
disordered models in Sec. \ref{disorder}. Let us remark that the very slow growth
of $R(t)$ in such systems as compared to the clean case (see Fig. \ref{fig_rt})
introduces severe limitations to the range 
of values of $R(t)$ than can be accessed.
In particular, choosing large values of 
$t_p$ (meaning large system sizes, see Eq.~(\ref{tp})) would prevent one to access
the region with $t\gg t_p$ (or, equivalently, $R_p(t)\gg L$),
the one in which we are mainly interested in, 
where the percolation structure has been fully
established. For this reason we will always work
with small or moderate values of $L$. In order to compare the clean system to the 
disordered ones in the same regimes, the same choice will be made for 
the pure system.

All the results contained in this Section are obtained
with periodic boundary conditions by averaging 
over a non-equilibrium ensemble with order $10^5-10^6$ of realisations. The system linear size is $L=512$ in units of the lattice spacing.

\vspace{2cm}

\subsection{Clean case} \label{pure}

\begin{figure}[h]
\centering
\rotatebox{0}{\resizebox{.95\textwidth}{!}{\includegraphics{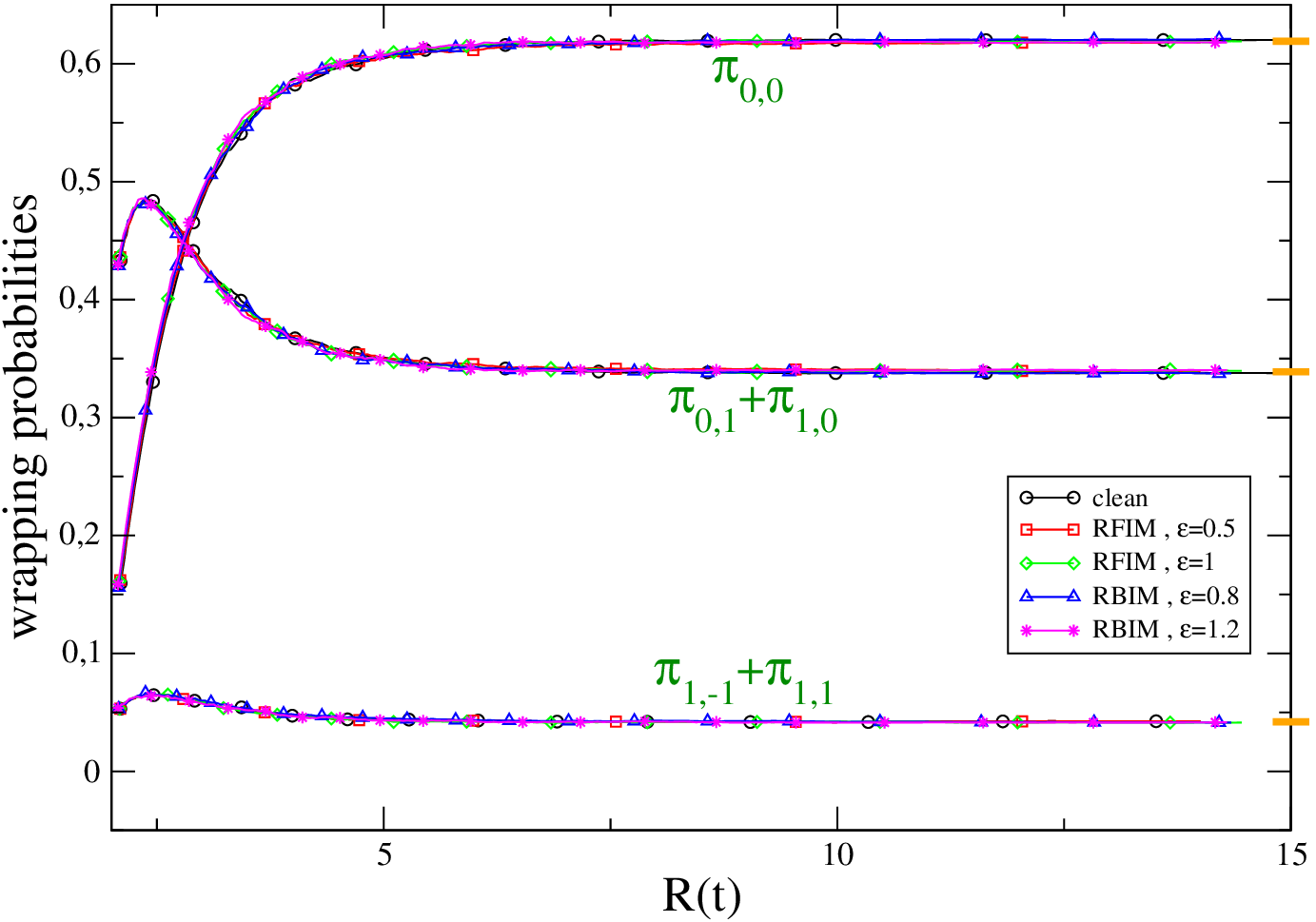}}}
\caption{The three wrapping probabilities $\pi_{0,0}$, $\pi_{0,1}+\pi_{1,0}$ and $\pi_{1,-1}+\pi_{1,1}$ 
  are plotted against $R(t)$ for a square lattice system with linear size
  $L=512$. 
Each set of superimposing curves represents one of these quantities
(as labelled in the figure) for different values of the disorder strength (see the key), including
the clean case. The exact values of random percolation are marked with a bold
(orange) segment on the extreme right.}
\label{fig_probcross}
\end{figure}

Let us consider a quench of the clean Ising model to $T_f=0$.
The wrapping cluster that develops around $t_p$ can cross the system
in different ways. The first possibility is to span the system from one side
to the other horizontally or vertically. We denote the probabilities of such configurations
as $\pi_{1,0}$ and $\pi_{0,1}$, respectively. Obviously, for a lattice with unit aspect ratio, $\pi_{0,1}=\pi_{1,0}$.
Another possibility is to have a domain traversing the sample in both the horizontal
and the vertical direction, the probability of which we indicate $\pi_{0,0}$.
Finally, clusters can percolate along one of the two diagonal directions with
equal probabilities $\pi_{1,-1}$ and $\pi_{-1,1}$. Since we operate with periodic boundary conditions,
other wrapping shapes -- winding the torus more than once -- are also possible
but these will not be considered in the following since they occur with an 
extremely small probability. The quantities mentioned above are exactly known
for two-dimensional critical percolation. They are~\cite{Pinson} 
\begin{eqnarray}
\pi_{0,1}+\pi_{1,0}&\simeq& 0.3388 \; , \nonumber \\
\pi_{0,0}&\simeq& 0.61908  \; , 
\label{spanpperc}
 \\
\pi_{1,-1}+\pi_{1,1} &\simeq& 0.04196 \; . \nonumber
\end{eqnarray}
Upon computing the wrapping probabilities defined above during the phase-ordering
process and plotting them against $R(t)$ (shown in Fig.~\ref{fig_rt})  
we find the curves shown in Fig. \ref{fig_probcross} 
(black curves with circles, which perfectly superimpose on
the others, represent the clean case at hand).
\vspace{2cm}

A first observation is
that any of these quantities, starting from zero immediately after the quench
($R(t)\simeq 0$) -- since
as already mentioned there cannot be any crossing in the initial state -- saturate
at long times (large $R(t)$) 
to values which are very precisely consistent with those given in Eq.~(\ref{spanpperc}) 
of critical percolation (bold orange segment on the far right).
This fact was first pointed out in \cite{Barros09,Olejarz12}.
The second observation is that all these probabilities attain their asymptotic
values around a certain $R(t)$ 
that can be used as a rough estimation of
$R(t_p)$. Concretely, it occurs at $R(t)\gtrsim  5$.  
Repeating the numerical experiment in systems with different sizes,
indeed, one can check that this determination of $R(t_p)$
increases with $L$, as it is expected after Eq.~(\ref{eq:Rp-Rd}).

\begin{figure}[h]
\centering
\rotatebox{0}{\resizebox{.95\textwidth}{!}{\includegraphics{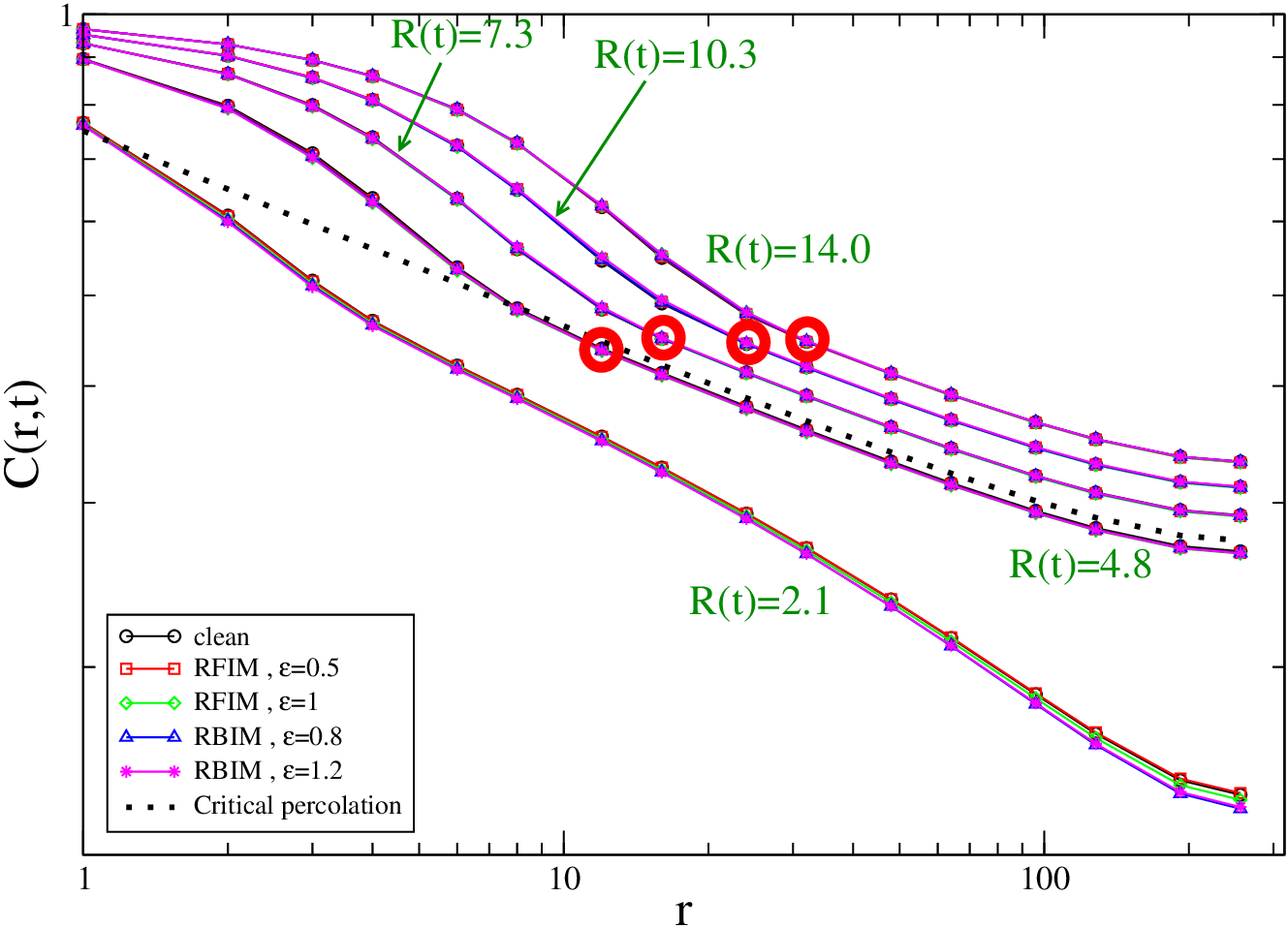}}}
\caption{The connectedness $C(r,t)$ is 
plotted against $r$. Each set of superimposing curves represents 
this quantity for a given value of $R(t)$ ($R(t)=2.1, 4.8, 7.3, 10.3, 14.0$, from
bottom to top, as indicated in the figure) for different
values of the disorder strength (see the key), including
the clean case (black curves with circles).
The same quantity computed numerically in a random percolation
square lattice at $p=p_c$ is 
represented by the dotted black curve, which agrees with the
exact form (\ref{connectperc}) apart from some finite size effects at large $r$.
The red circles correspond to the point where the percolative behavior
  starts to be observed.}
\label{fig_connect}
\end{figure}

The next quantity we consider is the pair connectedness defined in Eq.~(\ref{connect}).
In Fig. \ref{fig_connect} we plot this quantity against $r$ for different
values of $R(t)$, namely of time measured with the relevant clock (black curves with circles, which perfectly
superimpose with the others, represent the
clean case at hand). The area $S(t)=\sum _rC(r,t)$ below the curves 
increases as time elapses, because $S(t)$ is the probability that two points chosen at random in 
the system belong to the same cluster, and the number of domains decreases during
coarsening.
Regarding the form of each curve, after a first transient that we identify as
$t\lesssim t_p$, one observes the typical power-law behavior of critical
percolation, Eq.~(\ref{connectperc}) (dotted black line). This, however, is only true for values
of $r$ larger than a certain time-dependent value. This can be explained
as due to the fact that on scales smaller than $R(t)$ correlations set in and 
domains get compact. It is therefore natural to observe the percolative behavior
only at distances larger than $R(t)$. 
\vspace{2cm}

Let us now discuss the properties of $C$ more quantitatively, concerning in particular
its scaling behavior. According to the discussion around 
Eq.~(\ref{scal2par}) we expect a two-parameter scaling to hold also for the pair
connectivity, namely
\be
C(r,t)=c\left (\frac{r}{R(t)},\frac{L}{R_p(t)}\right ),
\label{scalC}
\ee
where $c$ is a scaling function.
In order to make the discussion simpler, let us start by discussing the scaling
properties for times much longer than the one -- $t_p$ -- where the percolating
structure sets in. This means that the second entry in Eq.~(\ref{scalC})
is small and one has
\be
C(r,t)\simeq c\left (\frac{r}{R(t)},0\right )={\cal C}\left (\frac{r}{R(t)}\right ),
\label{scalC1}
\ee
where ${\cal C}$ is a single-variable scaling function.
One can infer the form of the scaling function ${\cal C}(x)$ for large values of $x$
as follows. 
After $t_p$ the system has percolative properties for $r>R(t)$, and 
 it can be thought of as a percolation problem on a lattice with spacing $R(t)$,
for which Eq.~(\ref{connectperc}) must hold with the replacement $r_0 \to R(t)$,
namely
\be
C(r,t)\sim \left (\frac{r}{R(t)}\right )^{-2\Delta} \hspace{1cm} \mbox{for} \quad r\gg R(t).
\label{firstreason}
\ee
In order to match Eqs. (\ref{scalC1}) and (\ref{firstreason}) it must be 
\be
{\cal C}(x)\simeq x^{-2\Delta} \hspace{1cm} \mbox{for}\quad x\gg 1.
\label{beh1}
\ee
In the opposite situation of small distances $r\ll R(t)$ we are exploring the properties of
the domains well inside their correlated region, where there is no percolative
structure and the scaling function $c$ decays faster.
All the above holds for $t\gg t_p$ where the second entry in the scaling function
of Eq.~(\ref{scalC}) is very small. In the opposite situation $t\ll t_p$ 
(but $t$ larger than the microscopic time $t_0$ when scaling sets in) one has
\be
C(r,t)\simeq c\left (\frac{r}{R(t)},\infty \right )=\Xi \left (\frac{r}{R(t)}\right ).
\label{scalC2}
\ee
For $r\gg R(t)$ the scaling function $\Xi$ is expected to be different from 
${\cal C}$, because these two functions describe two contrasting situations where the percolative structure is, respectively, absent or present.
For $r\ll R(t)$, instead, the scaling function $\Xi$ should be akin to ${\cal C}$,
because in any case the domain structure is shaped by correlations on these short
lengthscales.
 
\begin{figure}[h]
  \centering
\rotatebox{0}{\resizebox{.95\textwidth}{!}{\includegraphics{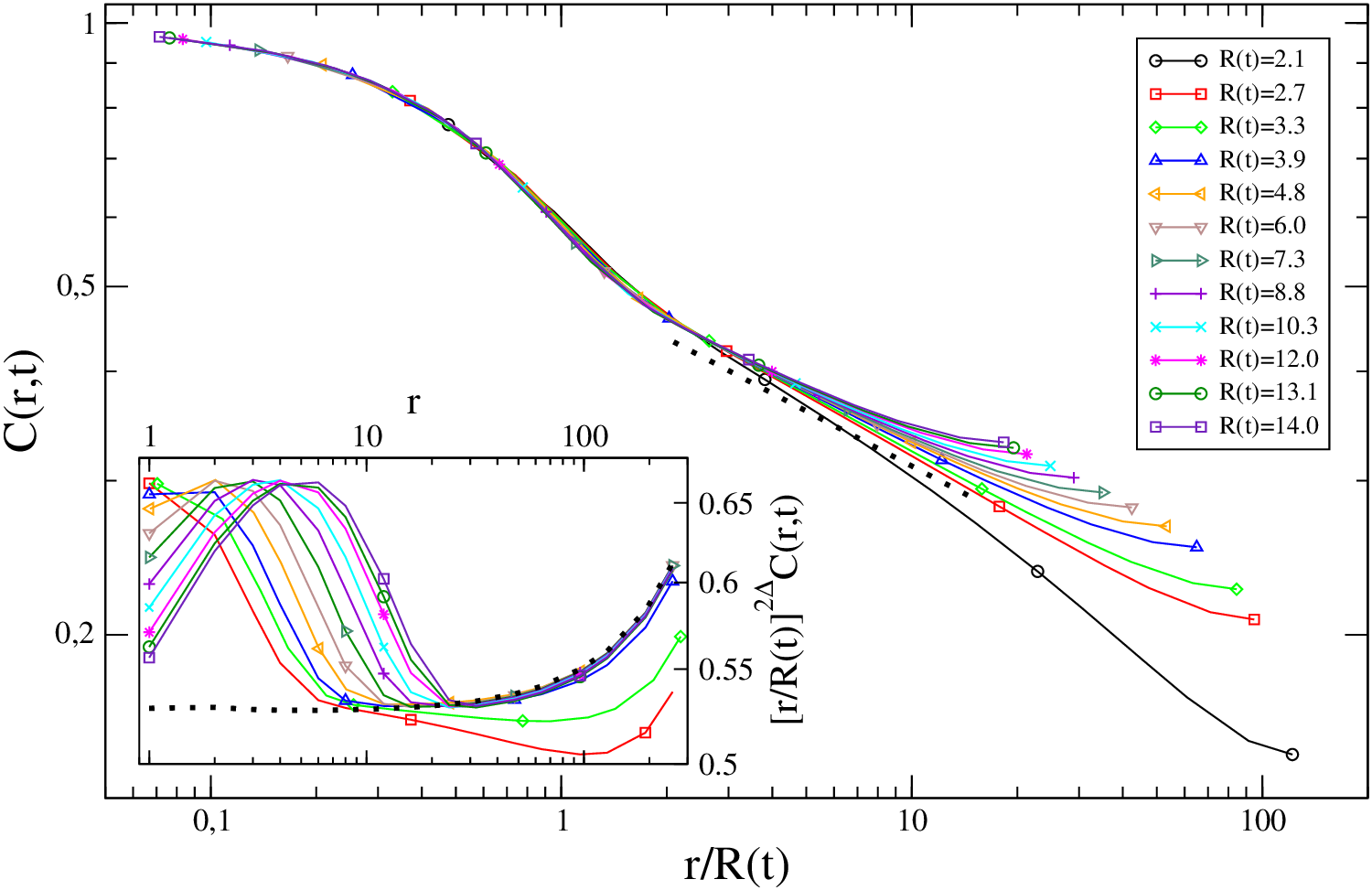}}}
\vspace{1cm}
\caption{The connectedness $C(r,t)$ against $r/R(t)$ in the clean case,
for different values of $R(t)$ (see the key) showing conventional scaling for short lengths
compared to $R(t)$.  The dotted black line is the behavior
(\ref{beh1}). In the inset, $(r/R(t))^{2\Delta} C(r,t)$ is plotted against $r$.
The dotted black line is the same curve as in Fig. \ref{fig_connect},
namely the connectedness function  computed numerically in a random percolation
square lattice at $p=p_c$.
The curves fail to scale on short lengths now but the plot demonstrates that the 
the long lengths behave as in critical percolation.}
\label{fig_scalingC}
\end{figure}

Let us submit these scaling ideas to the numerical test. 
In Fig. \ref{fig_scalingC} we plot $C(r,t)$ against
$r/R(t)$. 
We start by discussing the regime of sufficiently long times $t\gg t_p$.
Considering the previous rough estimate $R(t_p)\gtrsim 5$
obtained by inspection of the wrapping probabilities (see Fig.~\ref{fig_probcross}),
we can assume that such a regime can be sufficiently well represented by the curves with
$R(t)$ from $R(t)=4.8$ onward in 
Fig.~\ref{fig_scalingC}.
According to Eqs.~(\ref{scalC}) we should find collapse of such curves 
at different times on a unique mastercurve ${\cal C}$, with the behavior (\ref{beh1}).
The numerical data show indeed superposition, except in a region of large
$r$ (which moves towards smaller values of $r$ as time elapses)
where data collapse is lost due to finite size effects.
Notice also that, from $x\gtrsim x_{perc}=1.6$ the mastercurve behaves as in 
Eq.~(\ref{beh1}), as can be checked by comparing the numerical data 
with the dotted black line.
The departure from the behavior (\ref{beh1}) at very large $x$ is always
flanked by the failure of data collapse, a fact which strengthens the idea that
both these effects are due to finite size corrections.
Indeed, in the inset in the same figure we plot $(r/R(t))^{2\Delta} C(r,t)$ against
$r$ for all data in the main panel except the ones for $R(t)=2.1$.
The data collapse onto a master curve that is flat at not so large $r$ and then 
bends upwards due to finite-size corrections that can be captured by a correcting
factor $f(r/L)$ to be added as a factor to 
the scaling law (\ref{firstreason}). In the inset we also show the data for actual
critical percolation that 
superimpose on the dynamic data thus confirming that the bending of the curves is not a dynamic 
effect but just conventional finite-size corrections.
Let us now move to the regime with $t\ll t_p$. The curves
with $R(t)=2.1$ and $R=2.7$ in Fig. \ref{fig_scalingC} are representative of such regime.
We find that these two lines do not collapse for all values of $x>x_{perc}$.
This is because, since we work with relatively small values of $L$ 
(the reasons for that having been explained at the beginning of Sec. \ref{numres})
the regime with $R(t)\ll L$ where Eq.~(\ref{scalC2}) holds is not fully reached in our simulations.
However it is interesting to observe that in the region $x<x_{perc}$
data collapse is obeyed on a mastercurve
$\Xi$ which is almost indistinguishable from ${\cal C}$,
as expected according to the discussion below Eq.~(\ref{scalC2}).
On the other hand in the region 
$x> x_{perc}$ a marked difference is observed between the curves 
with  $R\le 2.7$ and those
with  $R(t)\ge 4.8$, as due to the fact that the percolative structure is still
absent when the former ones are computed whereas it has been established later. 

Finally, let us comment on the fact that the quantity $C$, besides being very informative about the 
twofold properties of the system -- compact and correlated at small distances vs fractal and uncorrelated at large
distances -- is also very well suited to assess the role of the {\it extra} growing length $R_p(t)$ in determining
the scaling properties. Indeed, although in \cite{Blanchard14} indications that
Eq.~(\ref{scal2par}) reproduces the data for $G$ better than 
the simple form (\ref{scal}) were given, 
the deviations from Eq.~(\ref{scal}) were relatively
small and located in the region of large $r$ where $G$ becomes very small. Instead,
as Fig.~\ref{fig_scalingC} clearly shows, a simple scaling form for $C$ completely
fails in describing the data, due to the differences between ${\cal C}$ and 
$\Xi$ in the region $x>x_{perc}$. For this quantity the improvement of Eq.~(\ref{scalC}) over 
a single parameter scaling is conspicuous.

\begin{figure}[h]
\centering
\rotatebox{0}{\resizebox{.95\textwidth}{!}{\includegraphics{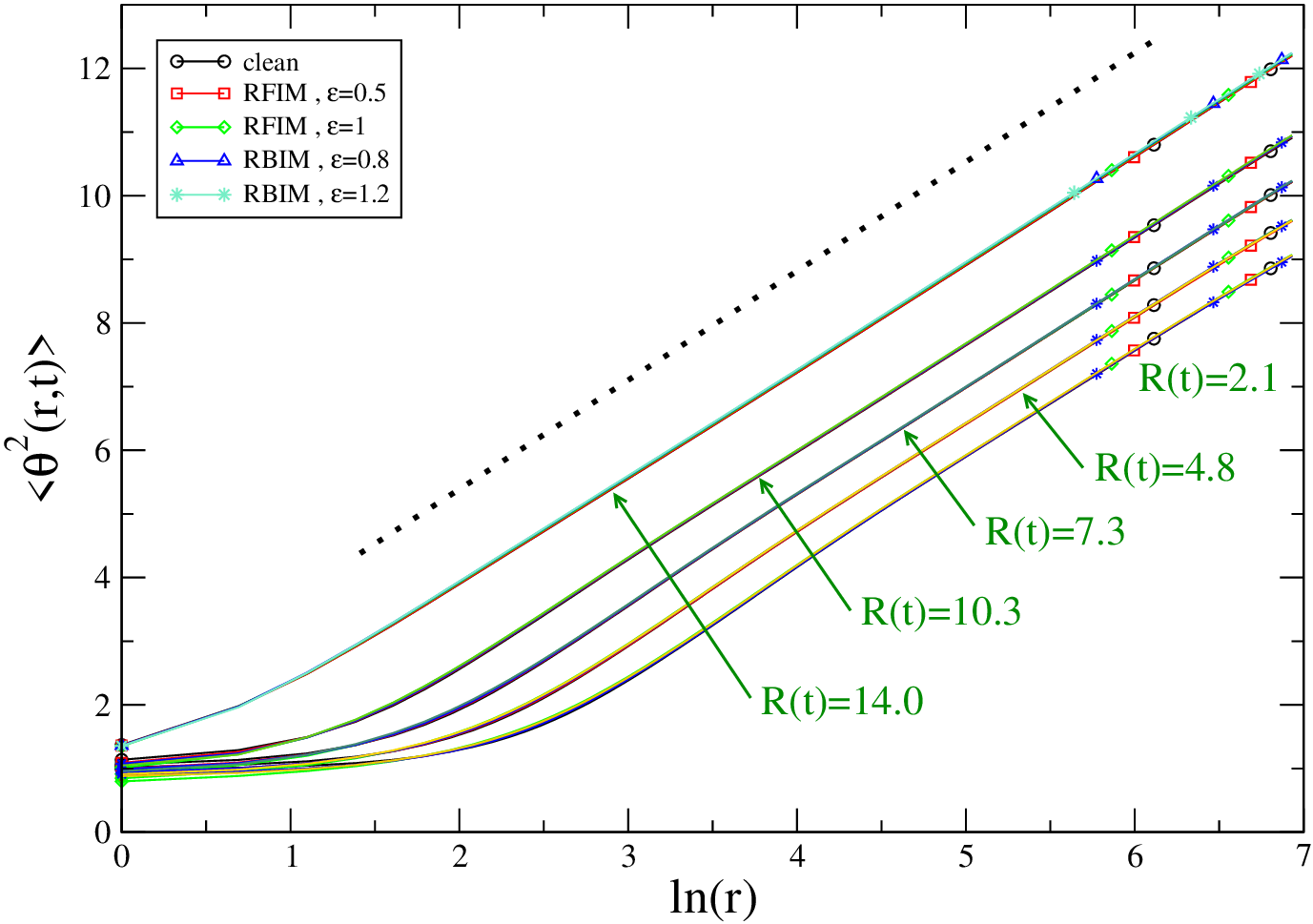}}}
\caption{The winding angle $\langle \theta^2(r,t)\rangle $ is 
plotted against $\ln r$. Each set of superimposing curve represents this quantity
for different values of the disorder strength (see the key), including
the clean case. The exact form (\ref{windexact}) of random percolation is 
the dotted black curve (the constant $a$ is arbitrarily fixed to be $a=2$).
}
\label{fig_wind}
\end{figure}

The last quantity that we will consider is the winding angle, that 
is plotted in Fig. \ref{fig_wind} (the clean case
corresponds to the black lines with circles, which are perfectly
superimposed to the others), for various choices of $R(t)$.

\begin{figure}[h]
\centering
\rotatebox{0}{\resizebox{.95\textwidth}{!}{\includegraphics{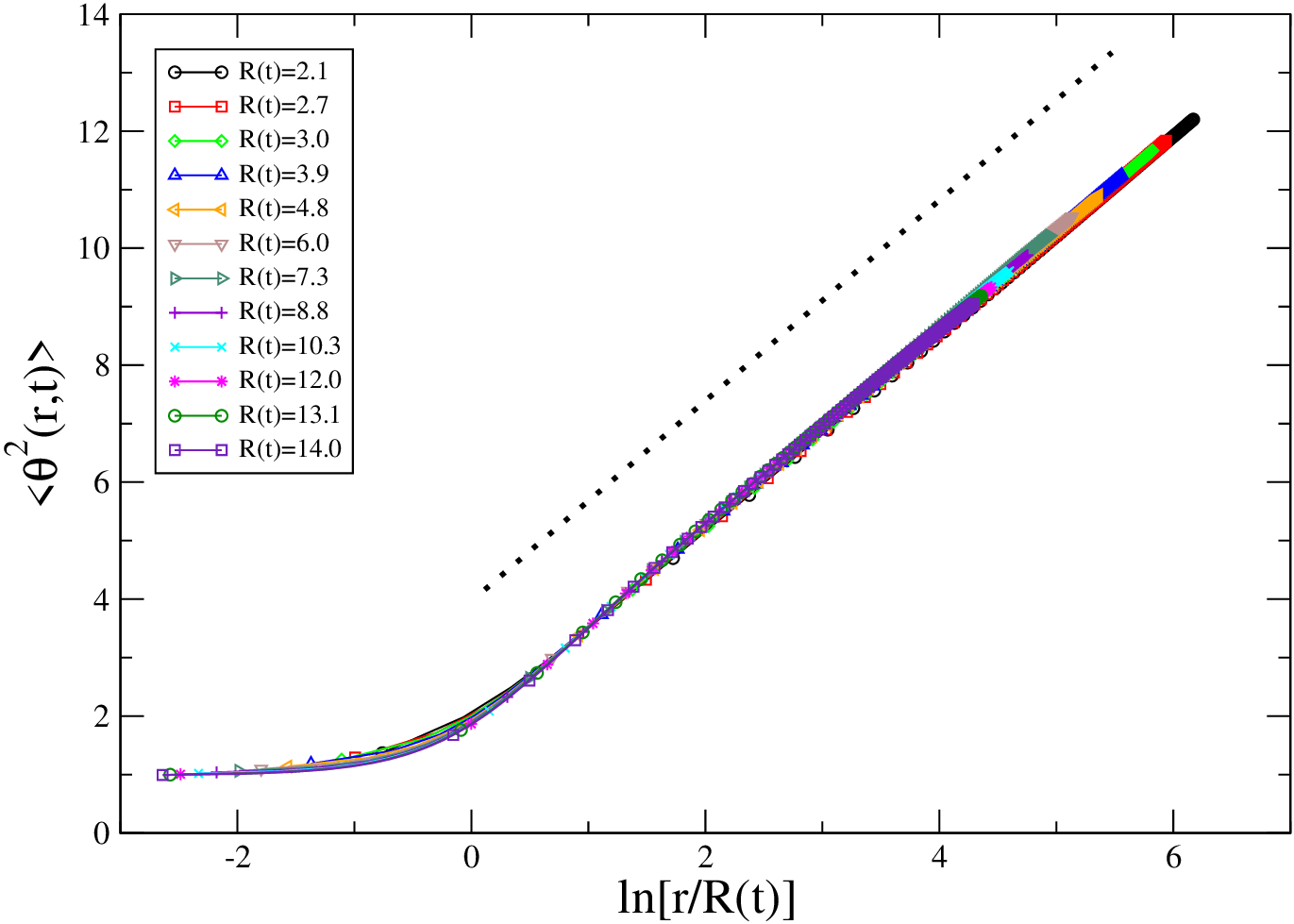}}}
\caption{The winding angle $\langle \theta^2(r,t) \rangle$ is 
  plotted against $\ln \left [r/R(t)\right ]$ for the pure case.
  The exact form (\ref{windexact}) of random percolation is 
the dotted black curve (the constant $a$ is arbitrarily fixed to be $a=4$).}
\label{fig_wind2}
\end{figure}

Following the discussion relative to the
connectivity, for $t>t_p$ we expect to observe the percolative behavior of Eq.~(\ref{windexact}),
with the replacement $r_0 \to R(t)$, namely~\cite{Duplantier,Wilson}
\be
\langle \theta ^2(r,t) \rangle=a+\frac{4k}{8+k} \ln \left ( \frac{r}{R(t)}\right ),
\label{windcoars}
\ee
for $r\gg R(t)$.  
If this is true we ought to find data collapse of the
curves for $\langle \theta ^2(r,t) \rangle$ at different times upon plotting them against $r/R(t)$,
if $r\gg R(t)$ is large enough and $R(t)\gg R(t_p)$ (similarly to what
previously observed for $C(r,t)$, see Fig. \ref{fig_scalingC}). In addition, the mastercurve
should be the one of Eq.~(\ref{windcoars}) for $r$ sufficiently
larger than $R(t)$. This kind of plot is
shown in Fig. \ref{fig_wind2}. Interestingly, not only we observe data collapse on the mastercurve
(\ref{windcoars}) when $t\gg t_p$, but this occurs with good precision also for a value  $R(t)=2.1$ which was shown in 
Fig. \ref{fig_scalingC} to be representative of a situation with $t< t_p$.
This implies that, for this particular quantity, a single parameter scaling 
\be
\langle \theta ^2(r,t) \rangle ={\cal T} \left ( \frac {r}{R(t)}\right ),
\label{scaltheta}
\ee
where ${\cal T}(x)$ is a scaling function, accounts
for the data. This does not contradict the general fact that 
-- due to the existence of 
the extra-length $R_p$ -- a two-parameter
scaling has to be expected. 
Indeed Eq.~(\ref{scaltheta}) is a particular case 
of a two-parameter form 
\be
\langle \theta ^2(r,t) \rangle =\Theta ^2\left (\frac{r}{R(t)},\frac{L}{R_p(t)}\right ),
\ee
where $\Theta ^2$ is a scaling function with a weak dependence on the second entry.

\subsection{Quenched disorder} \label{disorder}

In this Section we will discuss the effects of quenched disorder on the 
approach to percolation by considering the dynamics of the 
RFIM and the RBIM. We will investigate the properties of the same quantities
analyzed in the clean limit in Sec.~\ref{pure}.

Before starting this analysis let us briefly overview what is known about the coarsening 
behavior of weakly disordered models. When weak disorder is added, the
growth law $R(t)$ is markedly slowed down with respect to its growth in the pure case. 
This is observed in the RFIM and in the RBIM  as well.  
In particular, in the low temperature limit we are considering in this paper (i.e. $T\to 0$ with fixed $\epsilon$,
see the discussion around Eq.~(\ref{simplrates})), after a microscopic time, a regime 
in which an effective algebraic law, $R(t)\sim t^{1/\zeta(\epsilon)}$, is entered. The effective growth exponent,
starting from the value $\zeta (\epsilon =0)=2$ of the clean case, monotonously increases
with $\epsilon$~\cite{Paul07,Kolton}. This shows that in this stage $R(t)$ grows slower in the disordered 
models than in the clean case. All these features are quite neatly observed in Fig.~\ref{fig_rt},
where the behavior of $R(t)$  is shown.
At rather long times, for $t\gg t_{cross}$, a crossover occurs to an even slower growth, with
$R(t)$ increasing logarithmically, representing the asymptotic behavior~\cite{Fisher88,Kolton}. 
The simulations presented in this paper will never enter such an asymptotic stage, they will be 
restricted to $t< t_{cross}$. 

The slowing down of $R(t)$ reflects a fundamental difference in the coarsening mechanisms
at $\epsilon=0$ and $\epsilon \neq 0$. In the clean system phase-ordering 
proceeds by softening and flattening of interfaces, which is promoted by surface tension
and is a non-thermal effect. This means that activated processes are irrelevant and, as a consequence,
the kinetics has the same basic features at any $T_f$, including $T_f=0$
(which is indeed the case presented in the previous Section).
On the other hand, as soon as some disorder -- no matter how small -- is present, interfaces get
stuck in pinned configurations and the evolution can only be promoted by thermal activation.
These systems are frozen at $T_f=0$ (this is why we work at finite, although 
vanishing small, temperatures $T_f\to 0$). 

The basic result of this Section is that, despite the dynamics in the presence of disorder being
fundamentally different from the ones of the pure case, notwithstanding the fact 
that $R(t)$ increases in a 
radically different way, the percolative features observed in the clean case
occur with the same qualitative and quantitative modality in the presence of disorder.
Moreover, this occurs both for random fields and random bonds.

In order to prove the statement in the previous paragraph, let us start by discussing  the behaviour of the 
crossing probabilities introduced at the beginning of Sec.~\ref{pure}. 
These quantities are plotted in Fig.~\ref{fig_probcross}. 
For both the RFIM and the RBIM and for any strength of disorder 
considered parametrized by $\epsilon$
these probabilities are basically indistinguishable from those of the clean $\epsilon=0$ case.
This shows that the effect of disorder does not spoil the occurrence of the percolative structure, nor how and when 
(provided time is parametrized in terms of $R$) it is formed. As a byproduct, the independence of the crossing probabilities
on $\epsilon $ implies that Eq.~(\ref{zpzd}) is valid beyond the pure case with a 
unique value of the exponent $z_p/z_d$, at least for the non-conserved order parameter dynamics on the square lattice considered here.
Let us mention that, since $R(t)$ is strongly $\epsilon$-dependent, we should not find the nice
collapse of Fig.~\ref{fig_probcross} if we plotted against time, instead of against $R(t)$.
Equation~(\ref{zpzd}) would not look $\epsilon$-independent upon expressing $R_p$ as a function
of time either, namely in the form of Eq. (\ref{tp}). This shows that the typical size of correlated domains is the natural parametrization
of time in this problem. It will also help us conclude about the system size dependence of the 
percolation time $t_p$ without having to simulate different system sizes, as we will explain 
below. 

Let us move on to the connectedness function. A comparison between the clean case and the 
disordered ones is presented in Fig.~\ref{fig_connect}. In this figure one sees that 
all the disordered cases fall onto the clean case provided that times are chosen in order 
to have the same growing length. For instance, the curve for $t=103$ without disorder
is indistinguishable from the one at $t=1446$ for the RBIM with $\epsilon =1.2$, because 
at these times the size $R(t)$ of the domains in the two models is the same, $R(t)=14.0$ (last curve on the top in Fig. \ref{fig_connect}). 
Notice that collapse of the various curves is obtained for the RFIM and the 
RBIM, and for any strength $\epsilon$ of disorder. 
This confirms that the way in which the percolation structure develops, and the scaling 
properties associated to that, are those discussed in the previous Section regarding the clean case,
and are largely independent of the presence of disorder.

Finally, we arrive at the same conclusion by considering the winding angle.
This quantity is shown in Fig.~\ref{fig_wind} where 
the clean system and different disordered cases are shown. As in Fig.~\ref{fig_connect},
the comparison is made by choosing times so as to have the same $R(t)$.
Also in this case one observes an excellent superposition of all curves, further supporting
the conclusion that the presence of disorder does not modify the
percolative properties observed in the clean case. As a byproduct, this also mean that
a scaling plot like the one presented in Fig. \ref{fig_wind2} for the pure case
would look the same if data for the disordered models were used. This implies that
the scaling properties discussed in the previous Section for the winding angle apply
in the presence of quenched disorder as well. The same holds for the connectedness function.

As a final important point, let us discuss the role of the crossover time $t_{cross}$ where the growth law turns to a logarithmic form. As we said at the beginning of this Section,
our simulations do not even approach times of the order $t_{cross}$. However, the existence of this crossover time
is associated to a further characteristic length $\lambda (\epsilon)=R(t_{cross})$ which might be relevant to the scaling properties
discussed in Sec.~\ref{pure}. Indeed, considering the connectedness function for example, the dependence on 
$\lambda $ is expected to enter a scaling form in the following way   
\be
C(r,t)=c\left (\frac{r}{R(t)},\frac{L}{R_p(t)},\frac{R(t)}{\lambda(\epsilon)}
\right ),
\label{scalC2par}
\ee
where for simplicity we use the same symbol $c$ also for this new, three entry, scaling function.
Clearly, since in the range of times considered in our simulations we have $t\ll t_{cross}$, and consequently
$R(t)\ll \lambda(\epsilon)$, the last entry in the above equation is always around zero and can be neglected. 
This, in turn, makes our results independent of disorder $\epsilon $,
as we have already discussed, a fact that has been called 
{\it superuniversality} \cite{Fisher88}. Notice that this does not mean that disorder is ineffective, since  the growth law $R(t)$ changes 
dramatically due to the quenched randomness.
However, it is quite natural
to expect that such insensitivity of $c$ on $\epsilon $ could be spoiled if times where pushed to such late regimes as to give 
$R(t)\gtrsim \lambda(\epsilon)$. In this case the last entry in Eq.~(\ref{scalC2par}) could start, in principle, to play 
a role, introducing an $\epsilon $-dependence and spoiling the superuniversality we have shown to hold in the regime accessed in our simulations. The investigation 
of such a late regime requires a huge numerical effort that is beyond the scope of this paper and remains a challenge for future research.

\section{Conclusions} \label{concl}

In this paper we have investigated the relevance of percolative effects on the phase-ordering kinetics of the two-dimensional Ising model quenched from 
the disordered phase to a very low final temperature. We have considered the clean case as a benchmark, and two forms of quenched randomness, 
random bonds and random fields. The presence and the properties of the percolation cluster have been detected by inspection of quantities, such as the 
pair-connectedness, that represent an efficient tool to detect the percolative wrapping structure hidden in the patchwork of growing domains. 

The main finding of this paper is that the addition of weak quenched randomness, while sensibly changing the speed of the ordering process, does not impede 
the occurrence of 
percolation, nor it changes the way in which it sets in, even at a quantitative level. Indeed, we find that quantities such as the wrapping probabilities, the 
connectivity 
function and the winding angle behave in the same way with great accuracy for any choice of the disorder strength, including the clean case, once time has 
been measured in units of the size $R(t)$ of the ordered regions.      

All the above can be accounted for in a scaling framework where coarsening, percolation and disorder are associated to three characteristic lengths, $R$, $R_p$, and $\lambda $ respectively, and the interplay between them depends on how such lengths compare
between them and with the linear system size.

The results in~\cite{Sicilia08} and in this paper show that the relevance of percolation, 
which was previously pointed out for $2d$ clean systems, extends to the much less understood realm of disordered systems, making this issue of a quite general character. This not only opens the way to further studies on more general disordered systems (for instance, randomly diluted models where a more complex scaling structure has been recently observed \cite{Corberi13}), but also prompts the attention on a possible generalisation of analytical theories where the properties of phase-ordering are traced back to percolation effects, originally developed for clean systems, to the disordered cases.

 \vspace{1cm}
 
 \noindent
 {\bf Acknowkledgements}
 F. Corberi and F. Insalata thank the LPTHE Jussieu for hospitality during the preparation of this work. 
 L. F. Cugliandolo is a member of Institut Universitaire de France, 
and she thanks the KITP University of California at Santa Barbara for hospitality. This 
research was  supported in part by the National Science Foundation under 
Grant No. PHY11-25915.

\end{document}